\documentclass[article]{article}
\usepackage{color}
\usepackage{graphicx}
\usepackage{amsmath}
\usepackage{amssymb}
\usepackage{mathrsfs}
\usepackage[numbers,sort&compress]{natbib}
\usepackage{amsthm}
\usepackage{pgfplots}
\usepackage{float}
\usepackage{epstopdf}
\usepackage{xcolor}
\usepackage{array}
\usepackage{appendix}

\usepackage{booktabs} %调整表格线与上下内容的间隔
\usepackage{multirow}
\usepackage{subfigure}
\usepackage{geometry}

\numberwithin{equation}{section}
\allowdisplaybreaks[4]
\allowdisplaybreaks% 允许一个公式环境中的公式自动分页

\begin{document}

\baselineskip=18pt

\title{      Multiple-high-order pole solutions for  the   NLS  equation with quartic terms }
\author{Li-Li Wen$^1$, En-Gui Fan$^2$ and Yong Chen$^{1,3}$\thanks{\ Corresponding author and email address: ychen@sei.ecnu.edu.cn. } }
\footnotetext[1]{\  School of Mathematical Sciences, Shanghai Key Laboratory of Pure Mathematics and Mathematical Practice, East China Normal University, Shanghai, 200241, China.}
\footnotetext[2]{ \  School of Mathematical Sciences and Key Laboratory for Nonlinear Science, Fudan
University, Shanghai 200433, China.}

\footnotetext[3]{ \ College of Mathematics and Systems Science, Shandong University of Science and Technology, Qingdao, 266590, China.}

\date{}

\maketitle
\begin{abstract}
\baselineskip=15pt
 The aim of this article is to investigate the multiple-high-order pole solutions to  the focusing  NLS equation with quartic terms(QNLS) under the non-vanishing boundary conditions(NVBC) via the Riemann-Hilbert(RH) method.
  The determinant   formula of multiple-high-order pole soliton solutions for NVBC   is given. Further the double $1^{nd}$-order, mixed $2^{nd}$- and $1^{nd}$-order pole solutions are obtained.
\end{abstract}

\section{Introduction}
The NLS  equation can well describe the wave evolution of deep water and optical fibers \cite{ao2010,gp2006}. For some multiplicity of nonlinear phenomena in plasma waves, Bose-Einstein condensates and other physical phenomena, the NLS  equation is also a generic model\cite{hb2011,ep1961}. Nonetheless, several phenomena can not be described by low-order dispersion NLS equation. The NLS  equation with high-order dispersion is particularly important for describing the more complex phenomena.

In this article, we investigate the multiple-high-order pole soliton solution of focusing  QNLS equation\cite{ac2014}
\begin{equation}
iq_{t}+\frac{1}{2}q_{xx}+q|^{2}q+\gamma (q_{xxxx}+6|q|^{4}q+4q\bar{q}_{x}q_{x}+2q^{2}\bar{q}_{xx}+6\bar{q}q_{x}^{2}+8|q|^{2}q_{xx})=0. \label{nlsu}
\end{equation}
Soliton solutions  as one of the most important consequence of integrable systems, which can be constructed  by many different methods.    Recently, the Riemann-Hilbert(RH) method as another form of inverse scattering transform(IST) method is favorite to   study  the soliton solutions  with VBC \cite{yjk2010,wf},    as well as  NVBC \cite{gb2014,bp2006,wll,zy2020,zy2019,zys2020,mp2017}.
The $N$ simple poles and one $N$-order pole soliton of QNLS equation for VBC \cite{ln2020}. In the case of NVBC, the multiple-high-order pole solution not was given just investigated the pure one soliton in \cite{ln2020}.

The major contributions of this article are to study the QNLS equation  with NVBC via RH method.
We present the multiple-high-order pole  soliton solutions and construct the formula for soliton solution to be a determinant form.
For reflection-less and the scatter data $a(k)$ with multiple-high-order pole, the Talyor series on $a(k)$ and sectionally meromorphic matrix $M(x,t,z)$ are fully utilized. The advantage of this method is that the residue conditions of high-order poles are not the necessary conditions.

This work is organized as follows.  In  Section 2,   we   investigate IST method  of the QNLS equation with NVBC. And we show the analyticities, symmetries and asymptotic of Jost solution and scatter matrix.
In section 3, the  formulas   of single-high-order pole  and multiple-high-order pole solutions are shown.
We detailed given the double $1^{nd}$-order, mixed $2^{nd}$- and $1^{nd}$-order pole solutions.
\section{Inverse scattering transform for NVBC }
The QNLS equation admits the Lax pair
\begin{equation}
\psi_{x}=\mathfrak{h}\psi=(i\lambda\sigma_{3}+iQ)\psi,\quad\quad \psi_{t}=\mathfrak{l}\psi=
(-8i\gamma\sigma_{3}\lambda^{4}-8i\gamma Q\lambda^{3}+A_{2}\lambda^{2}+A_{1}\lambda+A_{0})\psi,\label{laxpair}
\end{equation}
where
\begin{subequations}
\begin{align}
&A_{2}=i(1+4\gamma Q^{2})\sigma_{3}-4\gamma\sigma_{3}Q_{x},\quad A_{1}=2\gamma[Q_{x},Q]\sigma_{3}+i(Q+4\gamma Q^{3}+2\gamma Q_{xx}),\nonumber\\
&A_{0}=-3i\gamma\sigma_{3}Q^{4}+i\gamma\sigma_{3}Q_{x}^{2}-i\gamma\sigma_{3}(QQ_{xx}+Q_{xx}Q)-\frac{i}{2}\sigma_{3}Q^{2}+\frac{1}{2}\sigma_{3}Q_{x}+6\gamma\sigma_{3}Q^{2}Q_{x}+\gamma\sigma_{3}Q_{xxx},\nonumber\\
&Q=\left(\begin{array}{cc}
   0& \overline{q}(x,t) \\
 q(x,t) & 0
 \end{array}
\right), \quad \sigma_{3}=\left(\begin{array}{cc}
  1 & 0 \\
  0 & -1
  \end{array}
\right).\nonumber
\end{align}
\end{subequations}
Considering the NVBC
$\lim\limits_{|x|\rightarrow\infty}q(x,t)=q_{\pm}=q_{0}e^{i\theta_{\pm}}$ where $|q_{\pm}|=q_{0}$. 
Naturally, $\mathfrak{h}_{\pm}=\lim\limits_{x\rightarrow\pm\infty}\mathfrak{h}, \quad\mathfrak{l}_{\pm}=\lim\limits_{x\rightarrow\pm\infty}\mathfrak{l}$, and $Q_{\pm}=\lim\limits_{x\rightarrow\pm\infty}Q$.
The eigenvalues of $\mathfrak{h}_{\pm}$ and $\mathfrak{l}_{\pm}$ are $\pm i\sqrt{\lambda^{2}+q_{0}^{2}}$ and
$\pm i\sqrt{\lambda^{2}+q_{0}^{2}}(\lambda-8\gamma\lambda^{3}+4\gamma\lambda q_{0}^{3})$, respectively.
Since the eigenvalues are doubly branched. One can transform the $\lambda$-plane to $z$-plane,
where 
$k=\frac{1}{2}(z+\frac{q_{0}^{2}}{z})$ and $\lambda=\frac{1}{2}(z-\frac{q_{0}^{2}}{z}).$

Now, we introduce the Jost solutions as the simultaneous solutions of Lax pair such that
\begin{equation}
\Psi_{\pm}(x,t,z)\sim\Omega_{\pm}(z)e^{i\theta(x,t,z)\sigma_{3}}, \quad z\in\Sigma, \quad x\rightarrow\pm\infty,
\end{equation}
where $\theta(x,t,z)=kx+k(\lambda-8\gamma\lambda^{3}+4\gamma\lambda q_{0}^{3})t$ and $\Sigma=\mathbb{R}\cup \mathcal{C}_{0}$. $\mathcal{C}_{0}$ is the circle of radius $q_{0}$ in the $z$-plane. For factoring the asymptotic exponential oscillations, we define a modified Jost solutions
\begin{equation}
\mu_{\pm}(x,t,z)=\Psi_{\pm}(x,t,z)e^{-i\theta(x,t,z)\sigma_{3}}.\label{mupsi}
\end{equation}
The modified Jost solutions $\mu_{\pm}(x,t,z)$ satisfy the asymptotic properties
$\lim\limits_{|x|\rightarrow\infty}{\mu}_{\pm}(x,t,z)=\Omega_{\pm}(z)$.
And the modified Jost solution admits the Volterra integral equations
\begin{equation}
\mu_{\pm}(x,t,z)=\Omega_{\pm}(z)+\int_{\pm\infty}^{x}\Omega_{\pm}(z)e^{ik(x-y)\hat{\sigma}_{3}}\Omega_{\pm}^{-1}\triangle Q_{\pm}\mu_{\pm}(y,t,z)\mathrm{d} y,\label{mu}
\end{equation}
where $\triangle{Q}_{\pm}=iQ-iQ_{\pm}$.
And one can obtain the analyticities of $\mu_{\pm,j}(x,t,z),(j=1,2)$: $\mu_{+,1}(x,t,z)$ and $\mu_{-,2}(x,t,z)$ analytic in $D_{+}=\{z\in\mathbb{C}\parallel(|z|^{2}-q_{0}^{2})\mathrm{Im}{z}>0\}$, $\mu_{-,1}(x,t,z)$ and $\mu_{+,2}(x,t,z)$ analytic in $D_{-}=\{z\in\mathbb{C}\parallel(|z|^{2}-q_{0}^{2})\mathrm{Im}{z}<0\}$.
The subscripts `1' and `2' identify the columns of matrix.

Since  $\Psi_{+}(x,t,z)$ and $\Psi_{-}(x,t,z)$ are the fundamental solutions of the spectral problem (\ref{laxpair}), so there exists a scatter matrix $S(z)=(s_{ij})_{2\times2}$ which satisfies the linear relationship
\begin{equation}
\mu_{+}(x,t,z)=\mu_{-}(x,t,z)e^{i\theta\sigma_{3}}S(z)e^{-i\theta\sigma_{3}},\quad z\in\Sigma\setminus\{\pm iq_{0}\},\label{S}
\end{equation}
Moreover, (\ref{S}) imply that $\det S(z)=1$.
Equation (\ref{S}) implies that $s_{11}(z)$ is analytic in $D_{+}$ and $s_{22}(z)$ is analytic in $D_{-}$.
  However, $s_{12}(z)$ and $s_{21}(z)$  are just continuous   on $\Sigma$.

For the QNLS equation with NVBC, there exist two kinds of symmetries for the Jost solution  $\Psi_{\pm}(z)$  and scattering matrix  $S(z)$ in $z$-plane:
\\
(1) The symmetries of up-half and low-half of $z$-plane ($z\leftrightsquigarrow\overline{z}$)
\begin{equation}
\Psi_{\pm}(z)=-\sigma\overline{\Psi_{\pm}(\overline{z})}\sigma,\quad \Psi_{\pm,j}=(-1)^{j-1}\sigma\overline{\Psi_{\pm,(3-j)}(\overline{z})},\quad S(z)=-\sigma\overline{S(\overline{z})}\sigma.\label{ssym1}
\end{equation}
(2) The symmetries of outside and inside of $\mathcal{C}_{0}$ ($z\leftrightsquigarrow-\frac{q_{0}^{2}}{z}$)
\begin{equation}
\Psi_{\pm}(z)=-\frac{1}{z}\Psi_{\pm}\big(-\frac{q_{0}^{2}}{z}\big)\sigma_{3}Q_{\pm},\quad\Psi_{\pm,j}(z)=(-1)^{j-1}\frac{1}{z}q_{\pm}\Psi_{\pm,(3-j)}\big(-\frac{q_{0}^{2}}{z}\big),\quad S(z)=Q_{-}^{-1}\sigma_{3}S\big(-\frac{q_{0}^{2}}{z}\big)\sigma_{3}Q_{+}.
 \label{ssym1}
\end{equation}
According to the above symmetries, the scattering matrix $S(z)$ can be rewritten as a new version
$S(z)=\left(
\begin{array}{cc}
   a(z)& -\overline{b(\overline{z})} \\
 b(z)& \overline{a(\overline{z})}
\end{array}\right)$.

Next, we consider the asymptotic of Jost solution and scattering matrix.
For convenient, we introduce the following notations
$A_{d}=\left(
\begin{array}{cc}
a_{11}&0\\
0&a_{22}
\end{array}
\right)$ and $A_{o}=\left(
\begin{array}{cc}
0&a_{12}\\
a_{21}&0
\end{array}
\right)$.
Consider the asymptotic expansion of $\mu_{\pm}(x,t,z)$
\begin{equation}
\mu_{\pm}(x,t,z)=\sum_{n=0}^{\infty}\mu_{\pm}^{(n)}(x,t,z),\label{mun}
\end{equation}
where
\begin{equation}
\mu_{\pm}^{(0)}(x,t,z)=\Omega_{\pm}(z)=I-\frac{1}{z}\sigma_{3}Q_{\pm},\quad\mu_{\pm}^{(n+1)}(x,t,z)=\int_{-\infty}^{x}\Omega_{\pm}(z)e^{ik(x-y)\hat{\sigma}_{3}}\Omega_{\pm}^{-1}(z)\triangle Q_{\pm}(y,t)\mu_{\pm}^{(n)}(y,t,z)\mathrm{d}y.\nonumber
\end{equation}
So
\begin{eqnarray}
\begin{split}
&\begin{split}\mu^{(n+1)}_{\pm,d}=&\int_{-\infty}^{x}\left(\Omega_{\pm,o}^{-1}\triangle Q_{\pm}\mu^{(n)}_{\pm,d}+\Omega_{\pm,d}^{-1}\triangle Q_{\pm}\mu^{(n)}_{\pm,o}\right)\mathrm{d}y\\
&+\int_{-\infty}^{x}\Omega_{\pm,o}e^{ik(x-y)\sigma_{3}}\left(\Omega_{\pm,d}^{-1}\triangle Q_{\pm}\mu^{(n)}_{\pm,d}+\Omega_{\pm,o}^{-1}\triangle Q_{\pm}\mu^{(n)}_{\pm,o}\right)e^{-ik(x-y)\sigma_{3}}\mathrm{d}y,\end{split}\\
&\begin{split}\mu^{(n+1)}_{\pm,o}=&\int_{-\infty}^{x}\left(\Omega_{\pm,o}\Omega_{\pm,o}^{-1}\triangle Q_{\pm}\mu^{(n)}_{\pm,d}+\Omega_{\pm,o}\Omega_{\pm,d}^{-1}\triangle Q_{\pm}\mu^{(n)}_{\pm,o}\right)\mathrm{d}y\\
&+\int_{-\infty}^{x}e^{ik(x-y)\sigma_{3}}\left(\Omega_{\pm,d}^{-1}\triangle Q_{\pm}\mu^{(n)}_{\pm,d}+\Omega_{\pm,o}^{-1}\triangle Q_{\pm}\mu^{(n)}_{\pm,o}\right)e^{-ik(x-y)\sigma_{3}}\mathrm{d}y.\end{split}
\end{split}\nonumber
\end{eqnarray}
If $z\rightarrow\pm\infty$, we note the fact that
$k=\frac{1}{2}(z+\frac{u_{0}^{2}}{z})$ and $\frac{1}{\omega}=1+\frac{u_{0}^{2}}{z^{2}}+\cdots$.
 And taking the integration by part for the last two terms of $\mu_{\pm,d}^{(1)}$ and $\mu_{\pm,o}^{(1)}$, one can derive the following results
\begin{equation}
\mu_{\pm,d}^{(1)}=O(\frac{\mu_{\pm,d}^{(0)}}{z})+O(\mu_{\pm,o}^{(0)})+O(\frac{\mu_{\pm,d}^{(0)}}{z^{2}})+O(\frac{\mu_{\pm,o}^{(0)}}{z^{3}})=O(\frac{1}{z}),\quad
\mu_{\pm,o}^{(1)}=O(\frac{\mu_{\pm,d}^{(0)}}{z^{2}})+O(\frac{\mu_{\pm,o}^{(0)}}{z})+O(\frac{\mu_{\pm,d}^{(0)}}{z})+O(\frac{\mu_{\pm,o}^{(0)}}{z^{2}})=O(\frac{1}{z}).\nonumber
\end{equation}
Iterate in the same method, one can obtain the asymptotic
\begin{equation}
\mu_{\pm,d}^{(2n)}=O(\frac{1}{z^{n}}),\quad\quad \mu_{\pm,o}^{(2n)}=O(\frac{1}{z^{n+1}}),\quad\quad \mu_{\pm,d}^{(2n+1)}=O(\frac{1}{z^{n+1}}),\quad\quad \mu_{\pm,o}^{(2n+1)}=O(\frac{1}{z^{n+1}}).\nonumber
\end{equation}
For $z\rightarrow0$, one can obtain the asymptotic in the similar way
\begin{equation}
\mu_{\pm,d}^{(2n)}=O(z^{n}),\quad\quad \mu_{\pm,o}^{(2n)}=O(z^{n-1}),\quad\quad \mu_{\pm,d}^{(2n+1)}=O(z^{n}),\quad\quad \mu_{\pm,o}^{(2n+1)}=O(z^{n}).\nonumber
\end{equation}
So we obtain the following asymptotic by some calculations for the equation (\ref{mun})
\begin{equation}
\mu_{\pm}(x,t,z)=I+O(\frac{1}{z}), \quad z\rightarrow\pm\infty,\quad\quad\mu_{\pm}(x,t,z)=-\frac{1}{z}\sigma_{3}Q_{\pm}+O(1),     \quad z\rightarrow 0.
\end{equation}
And one can derive the solution of QNLS equation
$q(x,t)=\lim\limits_{z\rightarrow\infty}z(\mu_{\pm})_{21}$.
By substituting the asymptotic of $\mu_{\pm}$ into (\ref{S}), one can obtain the asymptotic behaviors of the scattering matrix $S(z)$ 
\begin{equation}
S(z)=I+O(\frac{1}{z}),\quad z\rightarrow\pm\infty,\quad\quad S(z)=\frac{q_{+}}{q_{-}}I+O(z),\quad z\rightarrow 0.
\end{equation}

\section{The Riemann-Hilbert problem}
Based on the analyticity and asymptotic of  eigenfunctions  $\mu_\pm$ and $S(z)$, the solutions can be derived by RH problem. Firstly, we define a sectionally meromorphic matrix $M(x,t,z)$:
\begin{equation}
M^{+}(x,t,z)|_{z\in D_{+}}=\left(
\begin{array}{cc}
  \frac{\mu_{+,1}(x,t,z)}{a(z)}& \mu_{-,2}(x,t,z)
\end{array}\right),\quad M^{-}(x,t,z)|_{z\in D_{-}}=\left(
\begin{array}{cc}
  \mu_{-,1}(x,t,z)& \frac{\mu_{+,2}(x,t,z)}{\overline{a(\overline{z})}}
\end{array}\right).\nonumber
\end{equation}
\newtheorem{proposition}{\bf Proposition}[section]
\begin{proposition} The sectionally meromorphic matrix $M(x,t,z)$ satisfying the following RH problem:
\begin{itemize}
\item[$\bullet$] {Analyticity: $M(x,t,z)$ is  meromorphic  in $D_{+}\cup D_{-}$.}
\item[$\bullet$] {Jump condition:
$M^{+}(x,t,z)=M^{-}(x,t,z)\left(\begin{array}{cc}
   1-\rho\tilde{\rho}& -e^{2i\theta}\tilde{\rho} \\
e^{-2i\theta}\rho & 1
\end{array}\right), \quad z\in\Sigma$.}
\item[$\bullet$] {Asymptotic behavior:
$M(x,t,z)=I+O(\frac{1}{z}),~ z\rightarrow\infty,\quad M(x,t,z)=-\frac{1}{z}\sigma_{3}Q_{-}+O(1),~ z\rightarrow 0.$}
\end{itemize}
\end{proposition}
According to the asymptotic of $\mu_{\pm}$ and $S(z)$, the solution of QNLS can be rewritten as
$q(x,t)=\lim\limits_{z\rightarrow\infty}z(M(x,t,z))_{21}$.
For $a(z)$ with $N$ high order zeros, not only the residue conditions are useful but the coefficients of negative power should be considered. However, these coefficients are not straightforward derived. Moreover, if $z_{n}$ is the $N$-order zero point of $a(z)$, so is $-\frac{q_{0}^{2}}{\overline{z}_{n}}$. This situation can be equivalent to the zero points of $a(z)$ always paired in $D_{+}$.

We firstly consider single-high-order pole solutions.
We assume $a(z)$ with one $N$-order zero $z_{0}\in \{D_{+}\cap\mathrm{Im}{z}_{0}>0\}$. According to the symmetry properties of $S(z)$, one can derived $-\frac{q_{0}^{2}}{\overline{z}_{0}}$ also is the zero point of $a(z)$. For convenient, we make the following notations
$\{\nu_{1}\doteq z_{0}, \nu_{2}\doteq-\frac{q_{0}^{2}}{\overline{z}_{0}}, \overline{\nu}_{1}\doteq\overline{z}_{0}, \overline{\nu}_{2}\doteq-\frac{q_{0}^{2}}{z_{0}}\}$,
where $\nu_{1},\nu_{2}\in D_{+}$, and $\overline{\nu}_{1},\overline{\nu}_{2}\in D_{-}$.
So $a(z)$ can be expanded as the Taylor series
\begin{equation}
a(z)=a_{0}(z)\prod_{j=1}^{2}(z-\nu_{j})^{N},\quad j=1,2,\nonumber
\end{equation}
and $a_{0}(z)\neq0$ for all $z\in D_{+}$. Reflection coefficient $\rho(z)$ and $\overline{\rho(\overline{z})}$ with the Laurent expansion
\begin{equation}
\rho(z)=\rho_{j,0}(z)+\sum_{m_{j}=1}^{N}\frac{\rho_{j,m_{j}}}{(z-\nu_{j})^{m_{j}}},\quad\quad\overline{\rho(\overline{z})}=\overline{\rho_{j,0}(\overline{z})}+\sum_{m_{j}=1}^{N}\frac{\overline{\rho}_{j,m_{j}}}{(z-\overline{\nu}_{j})^{m_{j}}},\nonumber
\quad j=1,2,
\end{equation}
where
$\rho_{j,m_{j}}=\lim\limits_{z\rightarrow\nu_{j}}\frac{1}{(N-m_{j})!}\frac{\partial^{N-m_{j}}}{\partial z^{N-m_{j}}}[(z-\nu_{j})^{N}\rho(z)],\quad (m_{j}=1,2,\cdots,N),$
$\rho_{j,0}(z)$ is analytic for all $z\in D_{+}$.
According to the definition of $M(x,t,z)$, one can obtain that $z=\nu_{j}$ are the $N$-order poles of $M_{1}(x,t,z)$ and $z=\overline{\nu}_{j}$  are the $N$-order poles of $M_{2}(x,t,z)$. $M_{2}(x,t,z)$ analytic as $z=\nu_{j}$ and $M_{1}(x,t,z)$ analytic as $z=\overline{\nu}_{j}$. So we have the following expand
\begin{equation}
M_{21}(z)=\frac{q_{-}}{z}+\sum_{j=1}^{2}\sum_{p=1}^{N}\frac{G_{j,p}(x,t)}{(z-\nu_{j})^{p}},\quad M_{22}(z)=1+\sum_{j=1}^{2}\sum_{p=1}^{N}\frac{F_{j,p}(x,t)}{(z-\overline{\nu}_{j})^{p}}.
\end{equation}
$G_{j,p}(x,t)$ and $F_{j,p}(x,t)$ are undetermined. According to the analyticity one can get the Taylor expansion
\begin{equation}
e^{-2i\theta(z)}=\sum_{s_j=0}^{+\infty}f_{j,s_j}(x,t)(z-\nu_{j})^{s_j},\quad
M_{21}(z)=\sum_{s_j=0}^{+\infty}\zeta_{j,s_j}(x,t)(z-\overline{\nu}_{j})^{s_j},\quad
M_{22}(z)=\sum_{s_j=0}^{+\infty}\mu_{j,s_j}(x,t)(z-\nu_{j})^{s_j},\nonumber
\end{equation}
where
\begin{equation}
f_{j,s_j}(x,t)=\lim_{z\rightarrow\nu_{j}}\frac{1}{s_j!}\frac{\partial^{s_j}}{\partial z^{s_j}}e^{-2i\theta(z)},\quad \mu_{j,s_j}(x,t)=\lim_{z\rightarrow\nu_{j}}\frac{1}{s_j!}\frac{\partial^{s_j}}{\partial z^{s_j}}M_{22}(z),\quad
\zeta_{j,s_j}(x,t)=\lim_{z\rightarrow\overline{\nu}_{j}}\frac{1}{s_j!}\frac{\partial^{s_j}}{\partial z^{s_j}}M_{21}(z).\nonumber
\end{equation}
The Taylor expand of $e^{2i\theta(z)}$ also can be obtained. By considering the equation (\ref{S}) and the definition of $M(x,t,z)$, comparing the corresponding coefficients of $(z-\nu_{j})^{p}$ and $(z-\overline{\nu}_{j})^{-p}$. One can derived $G_{j,p}(x,t)$ and $F_{j,p}(x,t)$ as the form
\begin{equation}
F_{j,p}(x,t)=-\sum_{m_j=p}^{N}\sum_{s_j=0}^{m_j-p}\overline{\rho}_{j,m_{j}}\overline{f}_{j,m_j-p-s_j}(x,t)\zeta_{j,s_j}(x,t),\quad
G_{j,p}(x,t)=\sum_{m_j=p}^{N}\sum_{s_j=0}^{m_j-p}\rho_{j,m_{j}}f_{j,m_j-p-s_j}(x,t)\mu_{j,s_j}(x,t),\nonumber
\end{equation}
where $p=1,2,\cdots,N$. For $N=1$, $F_{j,p}(x,t)$ and $G_{j,p}(x,t)$ degenerate into the residue conditions. In addition, $\mu_{j,s_j}(x,t)$ and $\zeta_{j,s_j}(x,t)$ can be expressed as $F_{j,p}(x,t)$ and $G_{j,p}(x,t)$ via direct calculation
\begin{subequations}
\begin{align}
&\zeta_{j,s_j}(x,t)=\frac{(-1)^{s_j}q_{-}}{\overline{\nu}_{j}^{s_j+1}}+\sum_{p=1}^{N}\left(
\begin{array}{c}
  p+s_j-1\\
  s_j
\end{array}\right)\frac{(-1)^{s_j}G_{j,p}(x,t)}{(\overline{\nu}_{j}-\nu_{l})^{s_j+p}},\quad s_j=0,1,2,\cdots\nonumber\\
&\mu_{j,s_j}(x,t)=\begin{cases}
\begin{split}
&1+\sum_{p=1}^{N}\frac{F_{j,p}(x,t)}{(\nu_{j}-\overline{\nu}_{l})^{p}}, \quad s_j=0,\\
&\sum_{p=1}^{N}\left(
\begin{array}{c}
  p+s_j-1\\
  s_j
\end{array}\right)\frac{(-1)^{s_j}F_{j,p}(x,t)}{(\overline{\nu}_{j}-\nu_{l})^{s_j+p}}, \quad s_j=1,2,\cdots.\nonumber
\end{split}
\end{cases}
\end{align}
\end{subequations}
Then one can obtain the following system
\begin{eqnarray}
\begin{split}
&\begin{split}F_{j,p}(x,t)=&-\sum_{m_j=p}^{N}\sum_{s_j=0}^{m_j-p}\frac{(-1)^{s_j}\overline{\rho}_{j,m_{j}}\overline{f}_{j,m_j-p-s_j}(x,t)q_{-}}{\overline{\nu}_{j}^{s_j+1}}\\
&-\sum_{m_j=p}^{N}\sum_{s_j=0}^{m_j-p}\sum_{q=1}^{N}\left(
\begin{array}{c}
  q+s_j-1\\
  s_j
\end{array}\right)\frac{(-1)^{s_j}\overline{\rho}_{j,m_{j}}\overline{f}_{j,m_j-p-s_j}(x,t)G_{j,q}(x,t)}{(\overline{\nu}_{j}-\nu_{l})^{s_j+q}},\end{split}\\
&\begin{split}G_{j,p}(x,t)=&\sum_{m_j=p}^{N}\rho_{j,m_{j}}f_{j,m_j-p}(x,t)\\
&+\sum_{m_j=p}^{N}\sum_{s_j=0}^{m_j-p}\sum_{q=1}^{N}\left(
\begin{array}{c}
q+s_j-1\\
s_j
\end{array}
\right)\frac{(-1)^{s_j}\rho_{j,m_j}f_{j,m_j-p-s_j}(x,t)F_{j,q}(x,t)}{(\nu_{j}-\overline{\nu}_{l})^{s_j+q}}.\end{split}
\end{split}\label{FG}
\end{eqnarray}
For convenient, we introduce the following symbols ($s,p=1,2,\cdots,N$):
\begin{subequations}
\begin{align}
&|F\rangle=(\begin{array}{cccc}F_{1}&F_{2}&\cdots &F_{N}\end{array})^{T},\quad |G\rangle=(\begin{array}{cccc}G_{1}&G_{2}&\cdots& G_{N}\end{array})^{T},\quad|\beta\rangle=(\begin{array}{cc}
\beta_{1}&\beta_{2}
\end{array})^{T},\nonumber\\
&\beta_{j}=(\begin{array}{cccc}
\beta_{j,1}&\beta_{j,2}&\cdots&\beta_{j,N}\\
\end{array}),\quad|\eta\rangle=(\begin{array}{cc}
\eta_{1}&\eta_{2}
\end{array})^{T},\quad \eta_{j}=(\begin{array}{cccc}
\eta_{j,1}&\eta_{j,2}&\cdots&\eta_{j,N}\\
\end{array}),\nonumber\\
&\beta_{j,l}=-\sum_{m_j=l}^{N}\sum_{s_j=0}^{m_j-l}\frac{(-1)^{s_j}\overline{\rho}_{j,m_{j}}\overline{f}_{j,m_j-l-s_j}(x,t)q_{-}}{\overline{\nu}_{j}^{s_j+1}},\quad\eta_{j,l}(x,t)=\sum_{m_j=l}^{N}\rho_{j,m_j}f_{j,m_j-l}(x,t),\nonumber\\
&[\varpi_{j,l}]=\left([\varpi_{j,l}]_{p,q}\right)_{N\times N}=\left(-\sum_{m_{j}=p}^{N}\sum_{s_{j}=0}^{m_{j}-p}\left(\begin{array}{c}
q+s_{j}-1\\
s_{j}
\end{array}\right)
\frac{(-1)^{s_{j}}\overline{\rho}_{j,m_{j}}\overline{f}_{j,m_{j}-p-s_{j}}(x,t)}{(\overline{\nu}_{j}-\nu_{l})^{s_{j}+q}}
\right)_{N\times N}.\nonumber
\end{align}
\end{subequations}
and $\chi=\left(
\begin{array}{cc}
[\varpi_{11}]&[\varpi_{12}],\\
\left[\varpi_{21}\right]&[\varpi_{22}]
\end{array}
\right)$. So the equations (\ref{FG}) can be rewritten as
\begin{equation}
|F\rangle=|\beta\rangle+\chi|G\rangle,\quad\quad\quad |G\rangle=|\eta\rangle-\overline{\chi}|F\rangle.
\end{equation}
$|F\rangle$ and $|G\rangle$ are solved as
\begin{equation}
|F\rangle=\chi(I+\overline{\chi}\chi)^{-1}|\eta\rangle+(I-\chi(I+\overline{\chi}\chi)^{-1}\overline{\chi})|\beta\rangle,\quad\quad |G\rangle=(I+\overline{\chi}\chi)^{-1}(|\eta\rangle-\overline{\chi}|\beta\rangle),
\end{equation}
So the expansions of $M_{21}(x,t,\lambda)$ can be given as
\begin{equation}
M_{21}(x,t,z)=\frac{q_{-}}{z}+\sum_{j=1}^{2}\sum_{s=1}^{N}\frac{G_{j,s}(x,t)}{(z-\nu_{j})^{s}}=\frac{q_{-}}{z}+\frac{\det{[\mathrm{I}+\overline{\chi}\chi+(|\eta\rangle-\overline{\chi}|\beta\rangle)\langle\overline{Y(\overline{\lambda})}|]}}{\det{[\mathrm{I}+\overline{\chi}\chi]}}-1.
\end{equation}
where
\begin{equation}
\langle Y(z)|=\left(
\begin{array}{cccccccc}
\frac{1}{z-\overline{\nu}_{1}}&\frac{1}{(z-\overline{\nu}_{1})^{2}}&\cdots&\frac{1}{(z-\overline{\nu}_{1})^{N}}
&\frac{1}{z-\overline{\nu}_{2}}&\frac{1}{(z-\overline{\nu}_{2})^{2}}&\cdots&\frac{1}{(z-\overline{\nu}_{2})^{N}}
\end{array}
\right).\nonumber
\end{equation}
One can derive the solutions of equation QNLS with one $N$-order pole
\begin{equation}
q(x,t)=q_{-}+\left(\frac{\det{(\mathrm{I}+\overline{\chi}\chi+(|\eta\rangle-\overline{\chi}|\beta\rangle)\langle Y_{0}|)}}{\det{(\mathrm{I}+\overline{\chi}\chi)}}-1\right), \label{solution}
\end{equation}
where
\begin{subequations}
\begin{align}
&|\eta\rangle=(\begin{array}{cc}
\eta_{1}&\eta_{2}
\end{array})^{T},\quad|\beta\rangle=(\begin{array}{cc}\beta_{1}&\beta_{2}\end{array})^{T},\quad \langle Y_{0}|=(\begin{array}{cc}
Y_{1}^{0}&Y_{2}^{0}
\end{array}),\quad\eta_{j}=(\begin{array}{cccc}
\eta_{j,1}&\eta_{j,2}&\cdots&\eta_{j,N}\\
\end{array}),\nonumber\\
&\eta_{j,l}=\sum_{m_{j}=l}^{N}\rho_{j,m_{j}}f_{j,m_{j}-l}(x,t),\quad
\beta_{j}=(\begin{array}{cccc}
\beta_{j,1}&\beta_{j,2}&\cdots&\beta_{j,N}\\
\end{array}),\quad Y_{j}^{0}=(\begin{array}{cccc}
1&0&\cdots&0
\end{array})_{1\times N},\nonumber\\
&\beta_{j,l}=-\sum_{m_j=l}^{N}\sum_{s_j=0}^{m_j-l}\frac{(-1)^{s_j}\overline{\rho}_{j,m_{j}}\overline{f}_{j,m_j-l-s_j}(x,t)q_{-}}{\overline{\nu}_{j}^{s_j+1}},\quad
\chi=\left(
\begin{array}{cc}
[\varpi_{11}]&[\varpi_{12}]\\
\left[\varpi_{21}\right]&[\varpi_{22}]
\end{array}
\right),\nonumber
\end{align}
\end{subequations}
and $[\varpi_{j,l}](j,l=1,2)$ are $N\times N$ matrix
\begin{equation}
[\varpi_{j,l}]=\left([\varpi_{j,l}]_{p,q}\right)_{N\times N}=\left(-\sum_{m_{j}=p}^{N}\sum_{s_{j}=0}^{m_{j}-p}\left(\begin{array}{c}
q+s_{j}-1\\
s_{j}
\end{array}\right)
\frac{(-1)^{s_{j}}\overline{\rho}_{j,m_{j}}\overline{f}_{j,m_{j}-p-s_{j}}(x,t)}{(\overline{\nu}_{j}-\nu_{l})^{s_{j}+q}}
\right)_{N\times N}.\nonumber
\end{equation}

Next we consider the multiple-high-order pole solutions.
$a(z)$ with $N$ high-order zeros $z_{1}, z_{2},\cdots,z_{N}$, and the powers are $n_{1}, n_{2},\cdots,n_{N}$, respectively. Then $-\frac{q_{0}^{2}}{\overline{z}_{1}}, -\frac{q_{0}^{2}}{\overline{z}_{2}} ,\cdots, -\frac{q_{0}^{2}}{\overline{z}_{N}}$ also are the zeros of $a(z)$. And the powers are $n_{1}, n_{2},\cdots,n_{N}$, too. We also make the following notations
$\{\nu_{j}\doteq z_{i}, \nu_{j}|_{j=N+i}\doteq-\frac{q_{0}^{2}}{\overline{z}_{i}}, \overline{\nu}_{j}\doteq\overline{z}_{i}, \overline{\nu}_{j}|_{j=N+i}\doteq-\frac{q_{0}^{2}}{z_{i}}\}_{i=1}^{N}$.
So $a(z)$ can be expanded as
\begin{equation}
a(z)=a_{0}(z)(z-\nu_{1})^{n_{1}}\cdots(z-\nu_{N})^{n_{N}}(z-\nu_{N+1})^{n_{1}}\cdots(z-\nu_{2N})^{n_{N}},\nonumber
\end{equation}
Similar to the one high-order pole, $\rho(z)$ can be expanded as the Laurent series
\begin{equation}
\rho(z)=\rho_{j,0}(z)+\sum_{m_{j}=1}^{n_{j}}\frac{\rho_{j,m_{j}}}{(z-\nu_{j})^{m_{j}}},\quad\quad\overline{\rho(\overline{z})}=\overline{\rho_{j,0}(\overline{z})}+\sum_{m_{j}=1}^{n_{j}}\frac{\overline{\rho}_{j,m_{j}}}{(z-\overline{\nu}_{j})^{m_{j}}},\nonumber
\end{equation}
where
$\rho_{j,m_{j}}=\lim\limits_{z\rightarrow\nu_{j}}\frac{1}{(n_{j}-m_{j})!}\frac{\partial^{n_{j}-m_{j}}}{\partial(z-\nu_{j})^{n_{j}-m_{j}}}\left[(z-\nu_{j})^{n_{j}}\rho(z)\right]$,
and $\rho_{j,0}(z)$ is analytic for all $z\in D_{+}$ and $j=1,\cdots,2N$.
So, in the similar method, one can derive the soliton solution formula with multiple-high-order poles:
\newtheorem{thm}{\bf Theorem}[section]
\begin{thm}\label{nonth}
For the NVBC, if $a(z)$ with multiple-high-order zeros, then the soliton solutions of equation (\ref{nlsu}) is
\begin{equation}
q(x,t)=q_{-}+\left(\frac{\det{(\mathrm{I}+\overline{\chi}\chi+(|\eta\rangle-\overline{\chi}|\beta\rangle)\langle Y_{0}|)}}{\det{(\mathrm{I}+\overline{\chi}\chi)}}-1\right), \label{solution1}
\end{equation}
where \begin{subequations}
\begin{align}
&|\eta\rangle=(\begin{array}{cccc}
\eta_{1}&\eta_{2}&\cdots&\eta_{2N}
\end{array})^{T},\quad|\beta\rangle=(\begin{array}{cccc}\beta_{1}&\beta_{2}&\cdots&\beta_{2N}\end{array})^{T},\quad\langle Y_{0}|=(\begin{array}{cccc}
Y_{1}^{0}&Y_{2}^{0}&\cdots&Y_{2N}^{0}
\end{array}),\nonumber\\
&Y_{j}^{0}=(\begin{array}{cccc}
1&0&\cdots&0
\end{array})_{1\times n_{j}},\quad\eta_{j}=(\begin{array}{cccc}
\eta_{j,1}&\eta_{j,2}&\cdots&\eta_{j,n_{j}}
\end{array}),\quad \beta_{j}=(\begin{array}{cccc}\beta_{j,1}&\beta_{j,2}&\cdots&\beta_{j,n_j}\end{array}),\nonumber\\
&\eta_{j,l}=\sum_{m_{j}=l}^{n_{j}}\rho_{j,m_{j}}f_{j,m_{j}-l}(x,t),\quad\beta_{j,l}=-\sum_{m_j=l}^{n_{j}}\sum_{s_j=0}^{m_j-l}\frac{(-1)^{s_j}\overline{\rho}_{j,m_{j}}\overline{f}_{j,m_j-l-s_j}(x,t)q_{-}}{\overline{\nu}_{j}^{s_j+1}},\nonumber\\
&\chi=\left(
\begin{array}{cccc}
[\varpi_{11}]&[\varpi_{12}]&\cdots&[\varpi_{1(2N)}]\\
\left[\varpi_{21}\right]&[\varpi_{22}]&\cdots&[\varpi_{2(2N)}]\\
\vdots&\vdots&\cdots&\vdots\\
\left[\varpi_{(2N)1}\right]&[\varpi_{(2N)2}]&\cdots&[\varpi_{(2N)(2N)}]\\
\end{array}
\right),\nonumber
\end{align}
\end{subequations}
and $[\varpi_{j,l}](j,l=1,2,\cdots,2N)$ are $n_{j}\times n_{l}$ matrix
\begin{subequations}
\begin{align}
&[\varpi_{j,l}]=\left([\varpi_{j,l}]_{p,q}\right)_{n_{j}\times n_{l}}=\left(-\sum_{m_{j}=p}^{n_{j}}\sum_{s_{j}=0}^{m_{j}-p}\left(\begin{array}{c}
q+s_{j}-1\\
s_{j}
\end{array}\right)
\frac{(-1)^{s_{j}}\overline{\rho}_{j,m_{j}}\overline{f}_{j,m_{j}-p-s_{j}}(x,t)}{(\overline{\nu}_{j}-\nu_{l})^{s_{j}+q}}
\right)_{n_{j}\times n_{l}}.\nonumber
\end{align}
\end{subequations}
\end{thm}

$\mathbf{Double~1^{nd}-order ~  pole~ solution}$

Let $\nu_{1}$ and $\nu_{2}$ are the 1-order zeros point of $a(z)$, so is $\nu_{3}$ and $\nu_{4}$. The reflection coefficient $\rho(z)$ can be expand as Laurent series
$\rho(z)=\rho_{j,0}(z)+\frac{\rho_{j,1}}{z-\nu_{j}}$.
Now $\chi$ in (\ref{solution1}) is defined as
\begin{equation}
\chi=\left(\begin{array}{cccc}
[\varpi_{11}]&[\varpi_{12}]&[\varpi_{13}]&[\varpi_{14}]\\
\left[\varpi_{21}\right]&[\varpi_{22}]&[\varpi_{23}]&[\varpi_{24}]\\
\left[\varpi_{31}\right]&[\varpi_{32}]&[\varpi_{33}]&[\varpi_{34}]\\
\left[\varpi_{41}\right]&[\varpi_{42}]&[\varpi_{43}]&[\varpi_{44}]\\
\end{array}
\right)_{4\times4},\nonumber
\end{equation}
where $n_{1}=n_{2}=n_{2}=n_{4}=1,\quad j,l=1,2,3,4$
\begin{subequations}
\begin{align}
&|\eta\rangle=(\begin{array}{cccc}
\eta_{1}&\eta_{2}&\eta_{3}&\eta_{4}
\end{array})^{T},\quad |\beta\rangle=(\begin{array}{cccc}
\beta_{1}&\beta_{2}&\beta_{3}&\beta_{4}
\end{array})^{T},\quad\langle Y_{0}|=(\begin{array}{cccc}
1&1&1&1
\end{array}),\nonumber\\
&\beta_{j,l}=-\sum_{m_j=l}^{n_{j}}\sum_{s_j=0}^{m_j-l}\frac{(-1)^{s_j}\overline{\rho}_{j,m_{j}}\overline{f}_{j,m_j-l-s_j}(x,t)q_{-}}{\overline{\nu}_{j}^{s_j+1}},\quad \eta_{j,l}=\sum_{m_{j}=l}^{n_{j}}\rho_{j,m_{j}}f_{j,m_{j}-l}(x,t).\nonumber\\
&[\varpi_{j,l}]=[\varpi_{j,l}]_{p,q}=-\sum_{m_{j}=p}^{n_{j}}\sum_{s_{j}=0}^{m_{j}-p}\left(\begin{array}{c}
q+s_{j}-1\\
s_{j}
\end{array}\right)
\frac{(-1)^{s_{j}}\overline{\rho}_{j,m_{j}}\overline{f}_{j,m_{j}-p-s_{j}}(x,t)}{(\overline{\nu}_{j}-\nu_{l})^{s_{j}+q}},\nonumber
\end{align}
\end{subequations}

$\mathbf{Mixed~  2^{nd}-  ~and~  1^{nd}-order~ pole~ solution}$

Let $\nu_{1}$ and $\nu_{3}$ are 2-order zero point of $a(z)$, $\nu_{2}$ and $\nu_{4}$ are 1-order zeros of $a(z)$. The reflection coefficient $\rho(z)$ can be expand as Laurent series
$\rho(z)=\rho_{j,0}(z)+\sum_{m_{j}=1}^{n_{j}}\frac{\rho_{j,m_{j}}}{(z-\nu_{j})^{m_{j}}}$.
Now $\chi$ in (\ref{solution1}) is defined as
\begin{equation}
\chi=\left(\begin{array}{cccc}
[\varpi_{11}]&[\varpi_{12}]&[\varpi_{13}]&[\varpi_{14}]\\
\left[\varpi_{21}\right]&[\varpi_{22}]&[\varpi_{23}]&[\varpi_{24}]\\
\left[\varpi_{31}\right]&[\varpi_{32}]&[\varpi_{33}]&[\varpi_{34}]\\
\left[\varpi_{41}\right]&[\varpi_{42}]&[\varpi_{43}]&[\varpi_{44}]\\
\end{array}
\right)_{6\times6},\nonumber
\end{equation}
where $n_{1}=n_{3}=2,\quad n_{2}=n_{4}=1,\quad j,l=1,2,3,4$
\begin{subequations}
\begin{align}
&|\eta\rangle=(\begin{array}{cccc}
\eta_{1}&\eta_{2}&\eta_{3}&\eta_{4}
\end{array})^{T},\quad |\beta\rangle=(\begin{array}{cccc}
\beta_{1}&\beta_{2}&\beta_{3}&\beta_{4}
\end{array})^{T},\quad\langle Y_{0}|=(\begin{array}{cccccc}
1&0&1&1&0&1
\end{array}),\nonumber\\
&\beta_{j,l}=-\sum_{m_j=l}^{n_{j}}\sum_{s_j=0}^{m_j-l}\frac{(-1)^{s_j}\overline{\rho}_{j,m_{j}}\overline{f}_{j,m_j-l-s_j}(x,t)q_{-}}{\overline{\nu}_{j}^{s_j+1}},\quad \eta_{j,l}=\sum_{m_{j}=l}^{n_{j}}\rho_{j,m_{j}}f_{j,m_{j}-l}(x,t),\nonumber\\
&[\varpi_{j,l}]=[\varpi_{j,l}]_{p,q}=-\sum_{m_{j}=p}^{n_{j}}\sum_{s_{j}=0}^{m_{j}-p}\left(\begin{array}{c}
q+s_{j}-1\\
s_{j}
\end{array}\right)
\frac{(-1)^{s_{j}}\overline{\rho}_{j,m_{j}}\overline{f}_{j,m_{j}-p-s_{j}}(x,t)}{(\overline{\nu}_{j}-\nu_{l})^{s_{j}+q}}.\nonumber
\end{align}
\end{subequations}

\section*{Acknowledgments}

 This work is supported by the National Natural Science Foundation of China(No.12175069)
 and Science and Technology Commission of Shanghai Municipality (No.21JC1402500 and No.18dz2271000).

\end{document}